\documentclass[letterpaper,english]{article}
\usepackage[T1]{fontenc}
\usepackage[latin9]{inputenc}
\usepackage{babel}
\usepackage{array}
\usepackage{units}
\usepackage{multirow}
\usepackage{amstext}
\usepackage[unicode=true]
 {hyperref}
\usepackage{breakurl}

\makeatletter

\providecommand{\tabularnewline}{\\}

\newcommand{\lyxaddress}[1]{
\par {\raggedright #1
\vspace{1.4em}
\noindent\par}
}

\usepackage{endnotes}
\usepackage{geometry}
\renewcommand{\footnote}{\endnote}
 
\date{}
\usepackage{float}
\floatstyle{ruled}
\restylefloat{table}

\makeatother

\begin{document}

\title{On selective influences, marginal selectivity, \\and Bell/CHSH inequalities}

\author{Ehtibar N. Dzhafarov\textsuperscript{1} and Janne V. Kujala\textsuperscript{2}}

\maketitle

\lyxaddress{\begin{center}
\textsuperscript{1}Purdue University, Department of Psychological
Sciences, ehtibar@purdue.edu \\\textsuperscript{2}University of Jyv\"askyl\"a,
Department of Mathematical Information Technology, jvk@iki.fi
\par\end{center}}


\bigskip{}

\textsc{Running Head:} Selective influences and Bell/CHSH inequalities

\pagebreak{}
\begin{abstract}
The Bell/CHSH inequalities of quantum physics are identical with the
inequalities derived in mathematical psychology for the problem of
selective influences in cases involving two binary experimental factors
and two binary random variables recorded in response to them. The
following points are made regarding cognitive science applications:
(1) compliance of data with these inequalities is informative only
if the data satisfy the requirement known as marginal selectivity;
(2) both violations of marginal selectivity and violations of the
Bell/CHSH inequalities are interpretable as indicating that at least
one of the two responses is influenced by both experimental factors.\\
\end{abstract}
\pagebreak{}

\section{Introduction}

There are two independently motivated theoretical developments, one
in cognitive science (CS) and one in quantum mechanics (QM), that
lead to essentially identical mathematical formalisms. 

In CS, the issue is known as that of \emph{selective influences}:
given several experimental factors and several random variables recorded
in response to them, how can one determine whether each factor influences
only those response variables it is designed to influence? The issue
was introduced by Sternberg (1969), and for stochastically non-independent
responses, by Townsend (1984). A rigorous mathematical theory of selective
influences has been developed by Dzhafarov and colleagues (Dzhafarov,
2003; Dzhafarov \& Gluhovsky, 2006; Dzhafarov \& Kujala, 2010, 2012a-c,
in press; Dzhafarov, Schweickert, \& Sung, 2004; Kujala \& Dzhafarov,
2008).%
\footnote{For a mathematically accessible overview of this development, see
Chapter 10 of Schweickert, Fisher, and Sung (2012).%
} 

The ``parallel'' issue in QM is known as the investigation of the
(im)possibility of accounting for quantum entanglement in terms of
``hidden variables.'' It dates back to the papers by Einstein, Podolsky,
and Rosen (1935), Bohm and Aharonov (1957), and Bell (1964). In the
present paper we use the generalized form of Bell's inequalities developed
in Clauser and Horne (1974), often referred to as CHSH inequalities
(after Clauser, Horne, Shimony, \& Holt, 1969). Quantum entanglement
and Bell-type inequalities are mentioned in Wang, Busemeyer, Atmanspacher,
\& Pothos (in press) among the main conceptual frameworks of QM with
a potential of applicability to CS problems. Aerts, Gabora, and Sozzo
(in press) present an example of such an application, and our paper
is essentially an extended commentary on it.%
\footnote{For a detailed overview of the QM-SC parallels related to Bell-type
inequalities, see Dzhafarov and Kujala (2012a-b).%
} 

Our discussion is confined to the simplest experimental design in
which: (1) there are two \emph{experimental factors} (in CS) or \emph{measurement
procedures} (in QM), denoted $\alpha$ and $\beta$ and varying on
two levels each; (2) for each of the four treatments (combinations
of factor levels) two \emph{response variables} (in CS) or \emph{measurement
results }(in QM)\emph{ }are recorded, denoted $A$ and $B$; and (3)
for each treatment these $A$ and $B$ are generally \emph{stochastically
dependent random variables}. 

In CS it is often the case that hypothetically or normatively (e.g.,
in accordance with the instructions given to the participants), $A$
is supposed to be a response to $\alpha$ only, and $B$ a response
to $\beta$ only. In other words, the \emph{influences} of factors
$\alpha,\beta$ on responses $A,B$ are hypothesized to be \emph{selective},
\begin{equation}
A\leftarrow\alpha,B\leftarrow\beta,\label{eq:selective}
\end{equation}
as opposed to at least one of $A,B$ being influenced by both $\alpha$
and $\beta$. Thus, in Aerts et al. (in press) experiments, the factors
are two requests: $\alpha=$ ``choose an animal'' and $\beta=$
``choose an animal sound.'' The first factor has two variants (levels):
one is $a=$ ``Horse or Bear?'', the other is $a'=$ ``Tiger or
Cat?'' Factor $\beta$ also has two levels: $b=$ ``Growls or Whinnies?''
and $b'=$ ``Snorts or Meows?'' The responses $A_{\alpha\beta}$
and $B_{\alpha\beta}$ are the two choices made in response to $\alpha$
and $\beta$. We can denote the values of $A_{\alpha\beta}$ by $+1$
and $-1$ according as the first or second of the alternatives listed
in $\alpha$ is chosen; and we assign $+1/-1$ to $B_{\alpha\beta}$
analogously. For instance, the response pair $\left(A_{ab'},B_{ab'}\right)$
denotes two choices made in response to treatment (Horse or Bear?,
Snorts or Meows?); and $\left(A_{ab'}=-1,B_{ab'}=+1\right)$ means
that the two choices are Bear, Snorts.%
\footnote{For other examples of selective influence problems in behavioral sciences,
see Dzhafarov (2003).%
}

In QM, an analogue of this experiment is the simplest Bohmian version
of the Einstein-Podolsky-Rosen paradigm (EPR/B), involving two spin-$\nicefrac{1}{2}$
entangled particles, ``Alice's'' and ``Bob's.'' The measurement
procedures (factors) $\alpha$ and $\beta$ here consist in setting
detectors for measuring spins along certain spatial directions, $\alpha=a$
or $a'$ for Alice's particle, and $\beta=b$ or $b'$ for Bob's.
The measurement results (responses) are spins along these directions,
with possible values, $+1=$ ``spin-up'' and $-1=$ ``spin-down.''
To emphasize the analogy, each level of $\alpha$ ($a$ or $a'$)
can be presented as the ``requests to choose'' between spin-up and
spin-down along the corresponding direction; and analogously for each
level of $\beta$. Thus, $\left(A_{ab'}=-1,B_{ab'}=+1\right)$ means
that Alice's detector is set for direction $a$ and the spin measures
down, while Bob's detector is set for direction $b'$ and the spin
is up.

In both CS and QM, by repeatedly using the four treatments $\left(\alpha,\beta\right)$
one estimates for each of them the joint probabilities $\Pr\left(A_{\alpha\beta}=i,B_{\alpha\beta}=j\right),$
where $i$ and $j$ stand for $+1$ or $-1$.

\section{Selective influences and marginal selectivity}

In the following the terms ``factor'' and ``response'' will be
used generically, to include both the CS and QM meanings. It is easy
to show that whatever the factors $\alpha,\beta$ and responses $A,B$,
one can always find a random variable $R$ and two functions $A'_{\alpha\beta}=f\left(R,\alpha,\beta\right)$
and $B'_{\alpha\beta}=g\left(R,\alpha,\beta\right)$, such that, for
any treatment $\left(\alpha,\beta\right)$, the joint distribution
of $\left(A_{\alpha\beta},B_{\alpha\beta}\right)$ is the same as
that of the pair of random variables $\left(A'_{\alpha\beta},B'_{\alpha\beta}\right)$.
In symbols, denoting ``is distributed as'' by $\sim$,
\begin{equation}
\left(\begin{array}{c}
A_{\alpha\beta}\\
B_{\alpha\beta}
\end{array}\right)\sim\left(\begin{array}{c}
A'_{\alpha\beta}=f\left(R,\alpha,\beta\right)\\
B'_{\alpha\beta}=g\left(R,\alpha,\beta\right)
\end{array}\right).\label{eq:supergeneral}
\end{equation}
In CS, $R$ can be interpreted as a \emph{common source of randomness}
for the responses, accounting for both their stochasticity and non-independence.
In the EPR/B paradigm of QM, $R$ represents all ``hidden variables''
responsible for the joint distributions of the spins.%
\footnote{It may appear that $R$ should be of enormous complexity, but in fact,
when dealing with two binary factors and two binary responses, it
never has to be more complex than a discrete random variable with
$2^{8}$ values. This follows from an extended version of the Joint
Distribution Criterion (Dzhafarov \& Kujala, 2012c).%
} It was proposed in Dzhafarov (2003) that the selectiveness (\ref{eq:selective})
simply means that, referring to (\ref{eq:supergeneral}), $\beta$
is a dummy argument in $f$ and $\alpha$ a dummy argument in $g$.
That is, (\ref{eq:supergeneral}) acquires the form
\begin{equation}
\left(\begin{array}{c}
A_{\alpha\beta}\\
B_{\alpha\beta}
\end{array}\right)\sim\left(\begin{array}{c}
A'_{\alpha}=f\left(R,\alpha\right)\\
B'_{\beta}=g\left(R,\beta\right)
\end{array}\right),\label{eq:representation}
\end{equation}
for some random variable $R$ and some functions $f,g$.%
\footnote{If such a representation exists, $R$ need not be more than a 16-valued
random variable. This follows from the Joint Distribution Criterion,
as formulated in Fine (1982). See Dzhafarov \& Kujala (2012a).%
} In the context of EPR/B, this hypothesis amounts to the possibility
of a classical (non-quantum) explanation for the joint distributions
of the spins.

An immediate consequence of (\ref{eq:representation}) is \emph{marginal
selectivity} (Dzhafarov, 2003; Townsend \& Schweickert, 1983): for
any $\alpha,\beta$, the distribution of $A$$_{\alpha\beta}$ does
not depend on $\beta$, nor the distribution of $B_{\alpha\beta}$
on $\alpha$. In other words, 
\begin{equation}
\begin{array}{c}
\begin{array}{l}
\Pr\left(A_{\alpha,\beta=b}=+1\right)=\Pr\left(A_{\alpha,\beta=b'}=+1\right)=\Pr\left(A'_{\alpha}=+1\right),\\
\Pr\left(B_{\alpha=a,\beta}=+1\right)=\Pr\left(B_{\alpha=a',\beta}=+1\right)=\Pr\left(B'_{\beta}=+1\right).
\end{array}\end{array}\label{eq:marg constraint}
\end{equation}
In QM marginal selectivity is known under other names, such as \emph{causal
communication constraint} (see Cereceda, 2000). It is trivially satisfied
if one assumes that the entangled particles are separated by a space-like
interval (i.e., either of them may precede the other in time from
the vantage point of an appropriately moving observer). In CS marginal
selectivity can sometimes be ensured by an appropriate choice of $A,B$
(Dzhafarov \& Kujala, 2012a, c), but generally (and in particular,
in applications like in Aerts et al., in press) it may very well be
violated.

\section{Selective influences and CHSH inequality}

The CHSH inequality is another consequence of (\ref{eq:representation}):
it can be written as
\begin{equation}
\Gamma\leq2,\label{eq:CHSH}
\end{equation}
where, with $\mathrm{E}$ denoting expectation, 
\begin{equation}
\Gamma=\max\left\{ \mathrm{\pm E}\left[A_{ab}B_{ab}\right]\pm\mathrm{E}\left[A_{ab'}B_{ab'}\right]\pm\mathrm{E}\left[A_{a'b}B_{a'b}\right]\pm\mathrm{\mathrm{E}}\left[A_{a'b'}B_{a'b'}\right]:\textnormal{ \# of }+\textnormal{ signs is odd}\right\} .
\end{equation}
This inequality follows from but does not imply (\ref{eq:representation}),
as we see in the imaginary situation depicted in Table \ref{tab:Joint-probabilities-of}.

\begin{minipage}[t]{1\columnwidth}%
-------------------------------------------Insert Table \ref{tab:Joint-probabilities-of}
about here-------------------------------------------%
\end{minipage}

It is, of course, also possible that marginal selectivity is satisfied
but $\Gamma$ exceeds 2. This is the focal fact in the EPR/B experiments
(e.g., Aspect, Grangier, \& Roger, 1982). In the pattern of probabilities
in Table \ref{tab:The-CHSH-inequality}, $\Gamma=4$ (which is its
largest theoretically possible value), violating not only (\ref{eq:CHSH})
but also its quantum version, with $2\sqrt{2}$ substituting for $2$
(Landau, 1987). 

\begin{minipage}[t]{1\columnwidth}%
-------------------------------------------Insert Table \ref{tab:The-CHSH-inequality}
about here-------------------------------------------%
\end{minipage}

\section{When marginal selectivity is violated}

We see that the two consequences of (\ref{eq:representation}), the
CHSH inequality and marginal selectivity, are logically independent.
Together, however, they form a \emph{criterion} (a necessary and sufficient
condition) for (\ref{eq:representation}). This was first proved by
Fine (1982), even if with (\ref{eq:marg constraint}) implied rather
than stipulated.%
\footnote{That it was implied is apparent from the notation used: Fine writes
all marginal probabilities as depending on one factor only. So if
marginal selectivity is violated, Fine's inequalities are simply inapplicable.
A proof with explicit derivation of (\ref{eq:marg constraint}) and
(\ref{eq:CHSH}) is obtained as a special case of the Linear Feasibility
Criterion described in Dzhafarov and Kujala (2012a).%
}

Consider now Table \ref{tab:Probability-estimates-from} that presents
the experimental results reported in Aerts et al. (in press). Marginal
selectivity here is violated in all four cases. Thus, 
\[
0.135=P\left(Cat,Growls\right)+P\left(Cat,Whinnies\right)\neq P\left(Cat,Snorts\right)+P\left(Cat,Meows\right)=0.766.
\]
The difference being both large and statistically significant, we
conclude that no representation (\ref{eq:supergeneral}) for this
experiment reduces to a representation (\ref{eq:representation}).
The CHSH inequality here is violated too, $\Gamma=2.420$, but this
does not add ramifications to the rejection of (\ref{eq:representation}). 

\begin{minipage}[t]{1\columnwidth}%
-------------------------------------------Insert Table \ref{tab:Probability-estimates-from}
about here-------------------------------------------%
\end{minipage}

If a violation of marginal selectivity was obtained in the EPR/B context,
it would require a revolutionary explanation, as this would contradict
special relativity: a measurement procedure $\alpha$ applied to Alice's
particle cannot affect measurement results on Bob's particle, provided
they are separated by a space-like interval. But in the context of
an experiment like in Aerts et al. (in press), the explanation is
both simple and plausible: the choice between animals is influenced
not only by animal options but also by sound options; and analogously
for the choice between sounds. One is more likely to choose Cat over
Tiger if one also faces the choice between Snorts and Meows than if
it is between Growls and Whinnies. In fact, the explanation in terms
of the lack of selectiveness would be the simplest one even if in
Aerts et al. (in press) a value of $\Gamma$ exceeding 2 was obtained
with marginal selectivity satisfied. But an interpretation of the
influences exerted by $\alpha$ on $B$ and/or by $\beta$ on $A$
in this case would be less straightforward, and an invocation of the
EPR/B analogy more elucidating.

\section{Conclusion}

The main points are summarized in the abstract. But perhaps this is
not the end of the story. There is a huge gap between representations
(\ref{eq:supergeneral}) and (\ref{eq:representation}), and a systematic
theory is needed to study \emph{intermediate cases}. We recently began
this study (Dzhafarov \& Kujala, 2012c), confining it, however, to
the cases with marginal selectivity satisfied. It is conceivable that
the situations where it is violated could be shown within the framework
of a general theory to be structurally different depending on the
value of $\Gamma$. This would impart a diagnostic value to findings
like those reported in Aerts et al. (in press).

\subsection*{Acknowledgments}

This work is supported by NSF grant SES-1155956. We thank Peter Bruza
and Jerome Busemeyer for stimulating correspondence pertaining to
this commentary.

\theendnotes

\makeatletter

\renewcommand\@biblabel[1]{}

\makeatother

\pagebreak{}

\begin{table}[h]
\caption{Joint probabilities of $A=+1/-1$ and $B=+1/-1$ at four treatments
$\left(\alpha,\beta\right)$.$^{\;\dagger}$ \label{tab:Joint-probabilities-of}}

\begin{centering}
\begin{tabular}{cc}
 & \tabularnewline
 & \tabularnewline
\multirow{2}{*}{$\; A$} & \tabularnewline
 & \tabularnewline
 & \tabularnewline
\end{tabular}%
\begin{tabular}{c|c|c|c}
\multicolumn{1}{c}{} & \multicolumn{2}{c}{$B$} & \tabularnewline
\cline{2-3} 
$\; a,b$ & $+1$ & $-1$ & \tabularnewline
\hline 
\multicolumn{1}{|c|}{$+1$} & .25 & .25 & \multicolumn{1}{c|}{.5}\tabularnewline
\hline 
\multicolumn{1}{|c|}{$-1$} & .25 & .25 & \multicolumn{1}{c|}{.5}\tabularnewline
\hline 
 & .5 & .5 & \tabularnewline
\cline{2-3} 
\end{tabular}$\qquad$$\qquad$%
\begin{tabular}{c|c|c|c}
\multicolumn{1}{c}{} & \multicolumn{2}{c}{$B$} & \tabularnewline
\cline{2-3} 
$\; a,b'$ & $+1$ & $-1$ & \tabularnewline
\hline 
\multicolumn{1}{|c|}{$+1$} & .25 & .5 & \multicolumn{1}{c|}{.75}\tabularnewline
\hline 
\multicolumn{1}{|c|}{$-1$} & .0 & .25 & \multicolumn{1}{c|}{.25}\tabularnewline
\hline 
 & .25 & .75 & \tabularnewline
\cline{2-3} 
\end{tabular}
\par\end{centering}

\medskip{}

\begin{centering}
\begin{tabular}{cc}
 & \tabularnewline
\multirow{2}{*}{$A$} & \tabularnewline
 & \tabularnewline
 & \tabularnewline
\end{tabular}%
\begin{tabular}{|c|c|c|c|}
\cline{2-3} 
\multicolumn{1}{c|}{$a',b$} & $+1$ & $-1$ & \multicolumn{1}{c}{}\tabularnewline
\hline 
$+1$ & .25 & .35 & .6\tabularnewline
\hline 
$-1$ & .15 & .25 & .4\tabularnewline
\hline 
\multicolumn{1}{c|}{} & .4 & .6 & \multicolumn{1}{c}{}\tabularnewline
\cline{2-3} 
\end{tabular}$\qquad$$\qquad$%
\begin{tabular}{|c|c|c|c|}
\cline{2-3} 
\multicolumn{1}{c|}{$a',b'$} & $+1$ & $-1$ & \multicolumn{1}{c}{}\tabularnewline
\hline 
$+1$ & .25 & .45 & .7\tabularnewline
\hline 
$-1$ & .05 & .25 & .3\tabularnewline
\hline 
\multicolumn{1}{c|}{} & .3 & .7 & \multicolumn{1}{c}{}\tabularnewline
\cline{2-3} 
\end{tabular}
\par\end{centering}

\medskip{}

$^{\dagger\;}$The CHSH inequality (\ref{eq:CHSH}) is satisfied ($\Gamma=0$),
but marginal selectivity (\ref{eq:marg constraint}) is violated:
e.g., $0.5=\Pr\left(B_{ab}=+1\right)\neq\Pr\left(B_{a'b}=+1\right)=0.4$.
This rules out a representation (\ref{eq:representation}).

\medskip{}
\end{table}

\pagebreak{}

\begin{table}[h]
\caption{Joint probabilities of $A=+1/-1$ and $B=+1/-1$ at four treatments
$\left(\alpha,\beta\right)$.$^{\;\dagger}$ \label{tab:The-CHSH-inequality}}

\begin{centering}
\begin{tabular}{cc}
 & \tabularnewline
 & \tabularnewline
\multirow{2}{*}{$A$} & \tabularnewline
 & \tabularnewline
 & \tabularnewline
\end{tabular}%
\begin{tabular}{c|c|c|c}
\multicolumn{1}{c}{} & \multicolumn{2}{c}{$B$} & \tabularnewline
\cline{2-3} 
$\; a,b$ & $+1$ & $-1$ & \tabularnewline
\hline 
\multicolumn{1}{|c|}{$+1$} & .5 & 0 & \multicolumn{1}{c|}{.5}\tabularnewline
\hline 
\multicolumn{1}{|c|}{$-1$} & 0 & .5 & \multicolumn{1}{c|}{.5}\tabularnewline
\hline 
 & .5 & .5 & \tabularnewline
\cline{2-3} 
\end{tabular}$\qquad$$\qquad$%
\begin{tabular}{c|c|c|c}
\multicolumn{1}{c}{} & \multicolumn{2}{c}{$B$} & \tabularnewline
\cline{2-3} 
$\; a,b'$ & $+1$ & $-1$ & \tabularnewline
\hline 
\multicolumn{1}{|c|}{$+1$} & .5 & 0 & \multicolumn{1}{c|}{.5}\tabularnewline
\hline 
\multicolumn{1}{|c|}{$-1$} & 0 & .5 & \multicolumn{1}{c|}{.5}\tabularnewline
\hline 
 & .5 & .5 & \tabularnewline
\cline{2-3} 
\end{tabular}
\par\end{centering}

\medskip{}

\begin{centering}
\begin{tabular}{cc}
 & \tabularnewline
\multirow{2}{*}{$A$} & \tabularnewline
 & \tabularnewline
 & \tabularnewline
\end{tabular}%
\begin{tabular}{|c|c|c|c|}
\cline{2-3} 
\multicolumn{1}{c|}{$a',b$} & $+1$ & $-1$ & \multicolumn{1}{c}{}\tabularnewline
\hline 
$+1$ & .5 & 0 & .5\tabularnewline
\hline 
$-1$ & 0 & .5 & .5\tabularnewline
\hline 
\multicolumn{1}{c|}{} & .5 & .5 & \multicolumn{1}{c}{}\tabularnewline
\cline{2-3} 
\end{tabular}$\qquad$$\qquad$%
\begin{tabular}{|c|c|c|c|}
\cline{2-3} 
\multicolumn{1}{c|}{$a',b'$} & $+1$ & $-1$ & \multicolumn{1}{c}{}\tabularnewline
\hline 
$+1$ & 0 & .5 & .5\tabularnewline
\hline 
$-1$ & .5 & 0 & .5\tabularnewline
\hline 
\multicolumn{1}{c|}{} & .5 & .5 & \multicolumn{1}{c}{}\tabularnewline
\cline{2-3} 
\end{tabular}
\par\end{centering}

\medskip{}

$^{\dagger\;}$The CHSH inequality (\ref{eq:CHSH}) is violated ($\Gamma=4$)
with marginal selectivity satisfied. This rules out a representation
(\ref{eq:representation}).

\medskip{}
\end{table}

\pagebreak{}

\begin{table}[h]
\caption{Probability estimates from Table 1 of Aerts et al. (in press).$^{\;\dagger}$$^{\;\ddagger}$
\label{tab:Probability-estimates-from}}

\begin{centering}
\begin{tabular}{cc}
 & \tabularnewline
 & \tabularnewline
\multirow{2}{*}{$A$} & \tabularnewline
 & \tabularnewline
 & \tabularnewline
\end{tabular}%
\begin{tabular}{c|c|c|c}
\multicolumn{1}{c}{} & \multicolumn{2}{c}{$B$} & \tabularnewline
\cline{2-3} 
$\; a,b$ & Growls & Whinnies & \tabularnewline
\hline 
\multicolumn{1}{|c|}{Horse} & .049 & .630 & \multicolumn{1}{c|}{.679}\tabularnewline
\hline 
\multicolumn{1}{|c|}{Bear} & .259 & .062 & \multicolumn{1}{c|}{.321}\tabularnewline
\hline 
 & .308 & .692 & \tabularnewline
\cline{2-3} 
\end{tabular}$\qquad$$\qquad$%
\begin{tabular}{c|c|c|c}
\multicolumn{1}{c}{} & \multicolumn{2}{c}{$B$} & \tabularnewline
\cline{2-3} 
$a,b'$ & Snorts & Meows & \tabularnewline
\hline 
\multicolumn{1}{|c|}{Horse} & .593 & .025 & \multicolumn{1}{c|}{.618}\tabularnewline
\hline 
\multicolumn{1}{|c|}{Bear} & .296 & .086 & \multicolumn{1}{c|}{.382}\tabularnewline
\hline 
 & .889 & .111 & \tabularnewline
\cline{2-3} 
\end{tabular}
\par\end{centering}

\medskip{}

\begin{centering}
\begin{tabular}{cc}
 & \tabularnewline
\multirow{2}{*}{$A$} & \tabularnewline
 & \tabularnewline
 & \tabularnewline
\end{tabular}%
\begin{tabular}{|c|c|c|c|}
\cline{2-3} 
\multicolumn{1}{c|}{$a',b$} & Growls & Whinnies & \multicolumn{1}{c}{}\tabularnewline
\hline 
Tiger & .778 & .086 & .864\tabularnewline
\hline 
Cat & .086 & .049 & .135\tabularnewline
\hline 
\multicolumn{1}{c|}{} & .864 & .135 & \multicolumn{1}{c}{}\tabularnewline
\cline{2-3} 
\end{tabular}$\qquad$$\qquad$%
\begin{tabular}{|c|c|c|c|}
\cline{2-3} 
\multicolumn{1}{c|}{$a',b'$} & Snorts & Meows & \multicolumn{1}{c}{}\tabularnewline
\hline 
Tiger & .148 & .086 & .234\tabularnewline
\hline 
Cat & .099 & .667 & .766\tabularnewline
\hline 
\multicolumn{1}{c|}{} & .247 & .753 & \multicolumn{1}{c}{}\tabularnewline
\cline{2-3} 
\end{tabular}
\par\end{centering}

\medskip{}

$^{\dagger\;}$Marginal selectivity is violated, the CHSH inequality
is violated too ($\Gamma=2.420$).

\medskip{}

$^{\ddagger\;}$$n$/treatment = 81.

\medskip{}
\end{table}

\end{document}